\documentstyle[12pt,epsfig]{article}

\def\cO{{\cal O}} 
\def\cF{{\cal F}}

\begin{document}  
\vspace*{-2cm}  
\renewcommand{\thefootnote}{\fnsymbol{footnote}}  
\begin{flushright}  
hep-ph/9903507\\
DTP/98/100\\  
December 1998\\  
\end{flushright}  
\vskip 65pt  
\begin{center}  
{\Large \bf Renormalization Group Improved
Heavy Quark Production in Polarized $\gamma \gamma$ Collisions}\\
\vspace{1.2cm} 
{\bf  
Michael Melles${}^1$\footnote{Michael.Melles@durham.ac.uk} and  
W.~James~Stirling${}^{1,2}$\footnote{W.J.Stirling@durham.ac.uk}  
}\\  
\vspace{10pt}  
{\sf 1) Department of Physics, University of Durham,  
Durham DH1 3LE, U.K.\\  
  
2) Department of Mathematical Sciences, University of Durham,  
Durham DH1 3LE, U.K.}  
  
\vspace{70pt}  
\begin{abstract}
The experimental determination of the partial width $\Gamma ( H \longrightarrow
\gamma \gamma )$ of an intermediate mass Higgs is among the most important 
measurements at a future photon photon collider. Recently it was shown that
large  non-Sudakov as well as Sudakov double logarithmic (DL) corrections
can be summed to all orders in
the background process $\gamma \gamma\, (J_z=0) \longrightarrow q \overline{q}$.
It was found that positivity and stability of the cross section
was only restored at the four-loop level. One remaining
large source of uncertainty stems from
the fact that the scale of the strong coupling is unspecified within the
double logarithmic approximation. In this paper we include the 
leading and next-to-leading
order running coupling to all orders. We thus remove the inherent
scale uncertainty of both the exact one-loop and all-orders
DL result without encountering any Landau-pole singularities. 
The effect is significant and, for the non-Sudakov form factor, is
 found to correspond to an effective
scale of roughly $\alpha_s(9m_q^2)$.
\end{abstract}
\end{center}  
\vskip12pt

\setcounter{footnote}{0}  
\renewcommand{\thefootnote}{\arabic{footnote}}  
  
\vfill  
\clearpage  
\setcounter{page}{1}  
\pagestyle{plain} 
 
\section{Introduction} 
The model independent knowledge of the total Higgs width 
combined with respective branching ratios allows an experimental determination 
of various partial widths of the Higgs boson \cite{egn,ik}. It is therefore of fundamental
importance for a detailed understanding of the electroweak symmetry breaking
mechanism. For an intermediate mass Higgs, i.e. with a mass below the $W^+W^-$
threshold (in the MSSM mass window), a photon linear collider (PLC) offers so far the only possibility
of a direct and model independent measurement of $\Gamma_{tot}$. Using Compton backscattering 
\cite{g,g2,t} of
initially polarized electrons and positrons the partial width $\Gamma (
H \longrightarrow \gamma \gamma)$ can be determined with very
good precision at a PLC \cite{j}. This quantity
is of significant interest in its own right, since all charged massive particles
contribute in the loop which might therefore be sensitive to
 new physics. With the
respective branching ratio BR$(H \longrightarrow \gamma \gamma)$,
determined in measurements at the LHC (and possibly also the NLC), the total
width can be reconstructed.

The dominant non-Higgs background for this energy regime is $\gamma \gamma (J_z=0)
\longrightarrow q \overline{q}$ with $q=b,c$. While this background is suppressed by
$\frac{m_q^2}{s}$ at the Born level (unlike the $J_z=\pm 2$ channel),
higher-order QCD radiative (Bremsstrahlung) corrections in principle remove
this suppression \cite{bkos}. In addition, large virtual non-Sudakov double 
$\log(s/m_q^2)$ logarithms (DL)
are present which at one loop can even lead to a negative cross section \cite{jt,jt2}.
In Ref.~\cite{ms} we presented explicit three-loop results in the DL
approximation which revealed a factorization of hard (non-Sudakov) and
soft (Sudakov) double logarithms for this process and led to the all-orders
resummation in the form of a confluent hypergeometric function $_2F_2$: 
\begin{eqnarray}
\sigma^{DL}_{{\rm virt}+{\rm soft}} &=& \sigma_{\rm Born}
\left\{ 1 +
{\cal F} \;\;
_2F_2 (1,1;2,\frac{3}{2}; \frac{1}{2}
{\cal F} ) +
2 \; {\cal F} \;\;
_2F_2 (1,1;2,\frac{3}{2}; \frac{C_A}{4 C_F}
{\cal F} ) \right\}^2 \nonumber \\
&& \exp \left( \frac{ \alpha_s C_F}{\pi} \left[ \log \frac{s}{m_q^2} \left(
\frac{1}{2} - \log \frac{s}{4 l_c^2} \right) + \log \frac{s}{4 l_c^2} -1 +
\frac{\pi^2}{3} \right] \right) \label{eq:vps}
\end{eqnarray}
where ${\cal F} = - C_F \frac{\alpha_s}{4 \pi} \log^2 \frac{s}{m^2}$
is the one-loop hard form factor. The Born cross section 
is given by
\begin{equation}
\frac{d \sigma_{\rm Born}} {d \cos \theta}
\left( \gamma + \gamma (J_z=0) \longrightarrow
q + \overline{q} \right) = \frac{12 \pi \alpha^2 Q_q^4}{s} \frac{
\beta \left( 1 - \beta^4 \right)}{ \left(1 - \beta^2 \cos^2 \theta \right)^2}
\label{eq:Bdcs}
\end{equation}
where $\beta=\sqrt{1-\frac{4 m_q^2}{s}}$ denotes the quark velocity and $Q_q$ 
the charge of the quark with mass $m_q$. $\alpha=\frac{1}{137}$ is the fine structure
constant and $\sqrt{s}$ the center-of-mass energy of the initial photons.

In Ref. \cite{ms2} it was pointed out that one needs to include at least 
{\it four} loops (at  the cross section level) of the non-Sudakov
 hard logarithms in order
to achieve positivity and stability. A major source of uncertainty remained in
the scale choice of the QCD coupling, two possible `natural' choices ---
$\alpha_s(s)$ and $\alpha_s(m_q^2)$  --- 
yielding very different numerical results. 

In this work we include the exact leading and next-to-leading 
order running coupling into the
derivation of both the {\it massive} Sudakov and the novel hard (non-Sudakov) 
form factors, thus removing a major
source of error in both the exact one-loop and all-orders DL
results. Our technique is based on the observation that in a particular loop
in a multi-loop ladder, the correct scale for the strong coupling is the 
characteristic transverse momentum flowing round the loop.
 Details are given in Section~\ref{sec:rg}. In Section~\ref{sec:nr}
we present numerical results including the full one-loop radiative corrections
based on Refs.~\cite{bkos} and \cite{jt} with the new renormalization group
improved form factor. Section~\ref{sec:con} contains our conclusions.

\section{Renormalization Group Improved Form Factors} \label{sec:rg}

\begin{center}
\begin{figure}
\centering
\epsfig{file=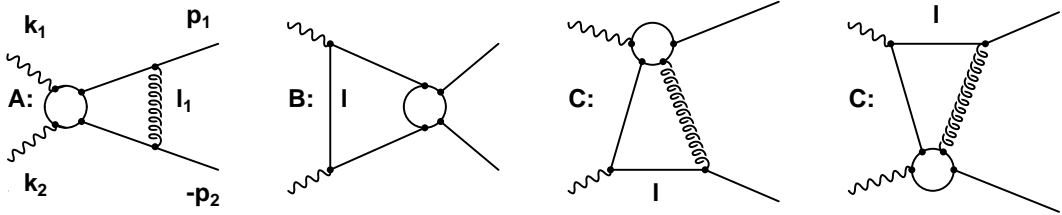,width=14cm}
\caption{The schematic one-loop soft (Sudakov,``A'') and hard (non-Sudakov,``B,C'')
topologies contributing to the DL form factor. These graphs are obtained from the
one-loop box diagram in the process $\gamma \gamma (J_z=0) \longrightarrow q 
\overline{q}$. The blob denotes a hard momentum flowing through the
omitted propagator relative to the soft momentum $l$ or $l_1$ in the DL phase 
space. For higher-order DL contributions, only corrections to these topologies
need to be taken into account.}
\label{fig:1l}
\end{figure}
\end{center}

In this section we will include the QCD running coupling into the DL form
factors given in Eq.~\ref{eq:vps}. A very important result incorporated in
this expression for the virtual plus soft real Bremsstrahlung cross section is
the {\it factorization} between Sudakov and non-Sudakov double logarithms
\cite{ms}. For our purposes here this means that it is sufficient to consider
the corrections for the basic topologies shown in Fig.~\ref{fig:1l} separately.
The complete RG improved result is then given by the same factorized 
structure.\footnote{In effect, introducing the running coupling softens the
contributions from the loops. It cannot promote sub-leading logarithms 
to the leading-logarithm level, and therefore cannot spoil the factorization.}

Within the DL approximation the scale at which to evaluate the QCD coupling
is unspecified, and this is therefore a major source of uncertainty in the
determination of the size of the $b\overline{b}$, $c\overline{c}$ background 
for intermediate mass Higgs production. In the following we will derive the
renormalization group effect of inserting a running coupling into each loop
evaluation using the exact one- and two-loop solution of the $\beta$-function. All
higher-order RG-terms will then be suppressed by $\frac{1}{\log^3 \frac{s}
{m^2}}$. 
We begin with the more familiar case of topology $A$ (see Fig.~\ref{fig:1l}).

\subsection{The Sudakov RG-Form Factor}

\begin{center}
\begin{figure}
\centering
\epsfig{file=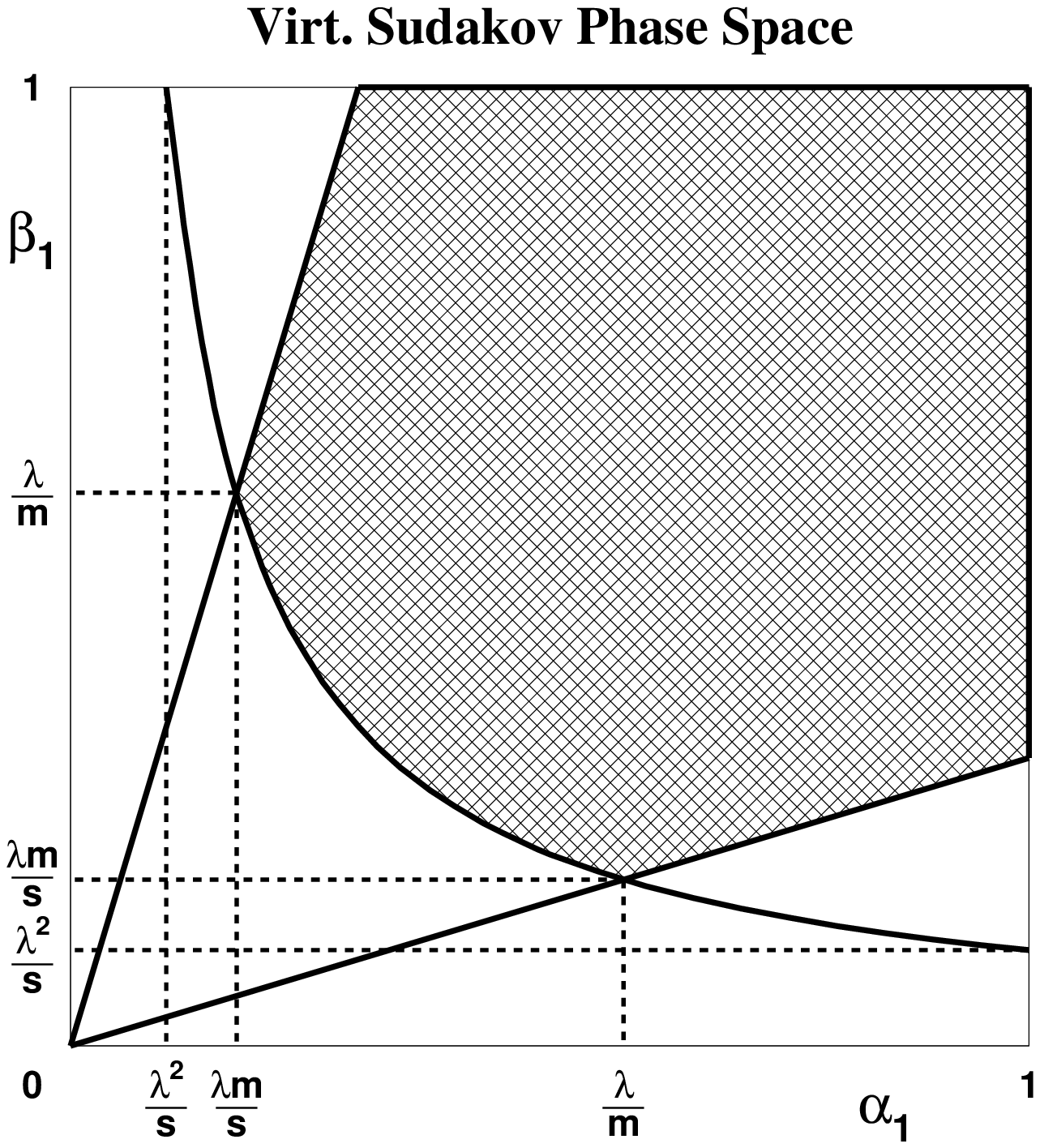,width=10cm}
\epsfig{file=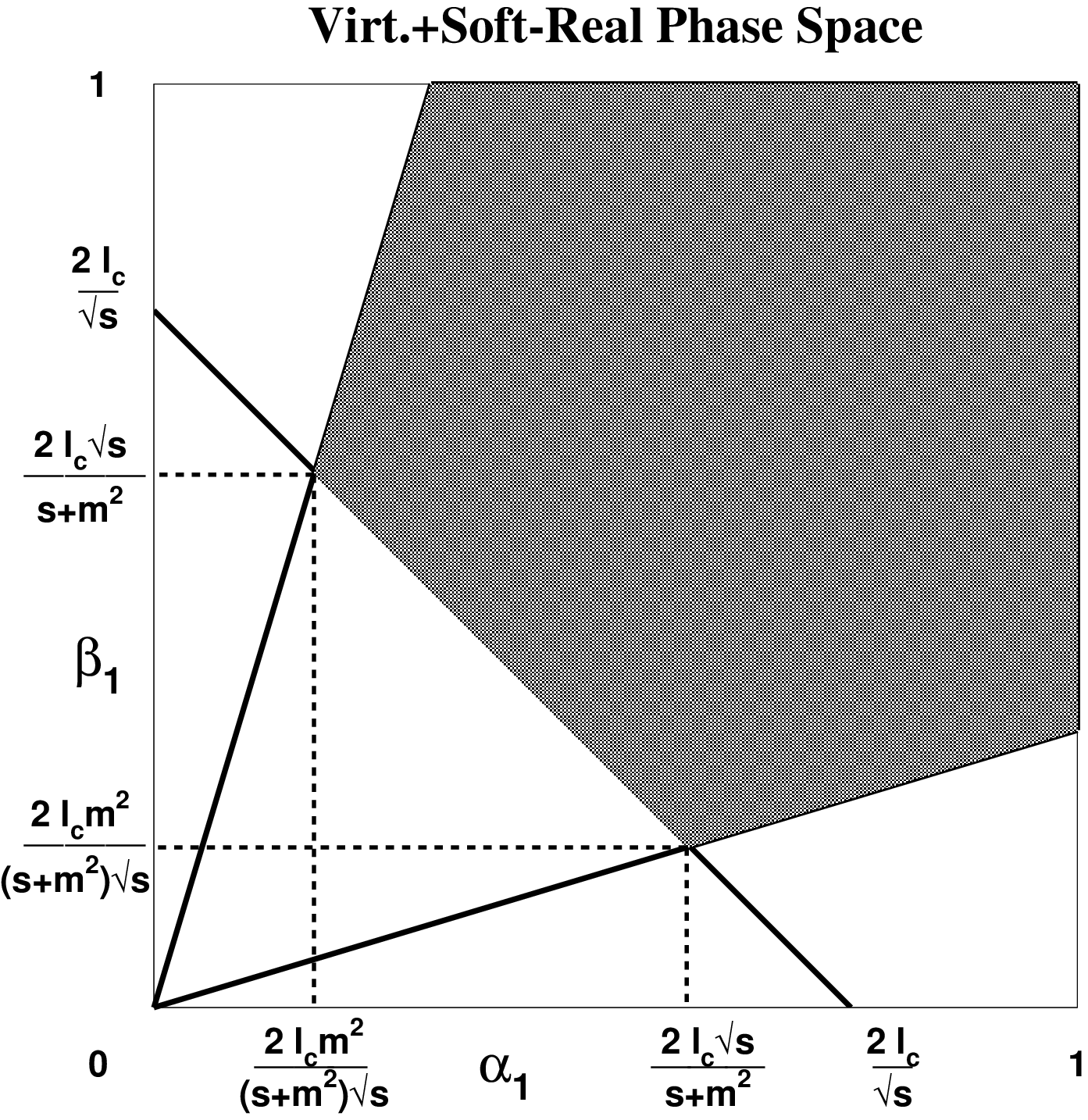,width=10cm}
\caption{The schematic DL-phase space for the massive virtual (upper diagram) and the
virtual plus soft real Sudakov form factor (lower diagram). 
The sum is gluon mass independent for
$\lambda < \frac{2 l_c m}{\sqrt{s}}$.}
\label{fig:sps}
\end{figure}
\end{center}
\begin{center}
\begin{figure}
\centering
\epsfig{file=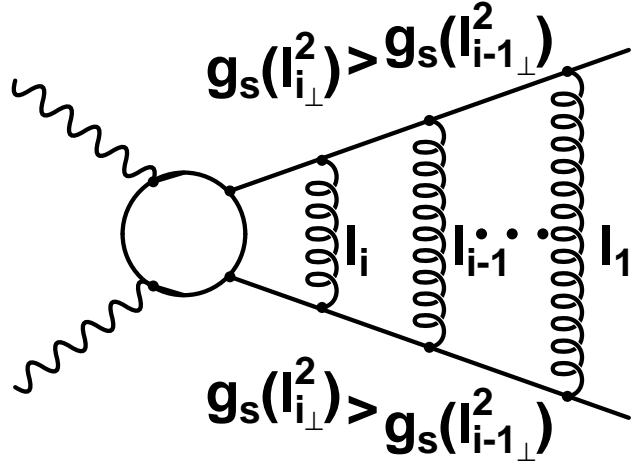,width=10cm}
\caption{A schematic Feynman diagram leading to the Sudakov
double logarithms in the process $\gamma \gamma (J_z=0) \longrightarrow q 
\overline{q}$ with $i$ gluon insertions. 
The blob denotes a hard momentum going through the omitted propagator
in the DL-phase space.
Crossed diagrams lead to a different
ordering of the Sudakov variables with all resulting $C_A$ terms canceling the
DL-contributions from three gluon insertions \cite{ms}.
The scale of the coupling $\alpha_s=\frac{g_s^2}{4
\pi}$ is indicated at the vertices and explicitly taken into 
account in this work.} 
\label{fig:SRG}
\end{figure}
\end{center}
In the derivation of the leading logarithmic corrections
in Ref.~\cite{ms} we used the familiar Sudakov technique \cite{Sud,gglf,fkm}
of decomposing loop momenta into components along 
external momenta, denoted by $\{\alpha, \beta\}$, and
those perpendicular to them, denoted by $l_\perp$.  
For massless fermions, the effective scale for Sudakov double logarithms 
for the coupling at each loop
is $\alpha_s ( {\bf l^2_\perp} )$, as was shown in
Refs.~\cite{b,ddt,cl} by direct comparison with explicit higher-order
calculations. For massive fermions the effective scale is also given by
${\bf l^2_\perp} \equiv - l^2_\perp > 0$ as the dominant 
double logarithmic phase
space is given by $\frac{m^2}{s} \ll \frac{{\bf l^2_\perp}}{s} \ll 1$ \cite{ms}
(on a formal DL level, setting $m=\lambda$ yields the massless Sudakov 
form factor).
We start by writing
\begin{equation}
\alpha_s ({\bf l^2_\perp})
 = \frac{\alpha_s(m^2)}{1+\beta_0 \frac{\alpha_s(m^2)
}{\pi} \log \frac{{\bf l^2_\perp}}{m^2}+\beta_1 \left( \frac{\alpha_s(m^2)}
{\pi} \right)^2 \log \frac{{\bf l^2_\perp}}{m^2}} 
\equiv \frac{\alpha_s(m^2)}{1+c \; \log \frac{{\bf l^2_\perp}}{m^2}} \label{eq:rc}
\end{equation}
where $\beta_0=\frac{11}{12} C_A -\frac{4}{12} T_F n_F$, 
$\beta_1 = \frac{17}{24} C_A^2 - \frac{5}{12} C_A T_F n_F- \frac{1}{4} 
C_F T_F n_F$ and for QCD we have
$C_A=3$, $C_F=\frac{4}{3}$ and $T_F=\frac{1}{2}$ as usual. 
Up to two loops the massless $\beta$-function is independent of the chosen
renormalization scheme and is gauge invariant in minimally subtracted schemes
to all orders \cite{c}. These features will also hold for the derived
renormalization group improved form factors below.

In case of the Sudakov DL's, the DL-phase space structure is very transparent 
in terms of the $\{ \alpha_1,\beta_1 \}$ parameters. This is shown in 
Fig.~\ref{fig:sps} for the virtual and virtual plus soft real Bremsstrahlung phase space.
We therefore use the on-shell condition ${\bf l^2_{1_\perp}}=s \alpha_1 \beta_1$, even 
though the running coupling will now depend on {\it two} integration variables.

\begin{center}
\begin{figure}
\centering
\epsfig{file=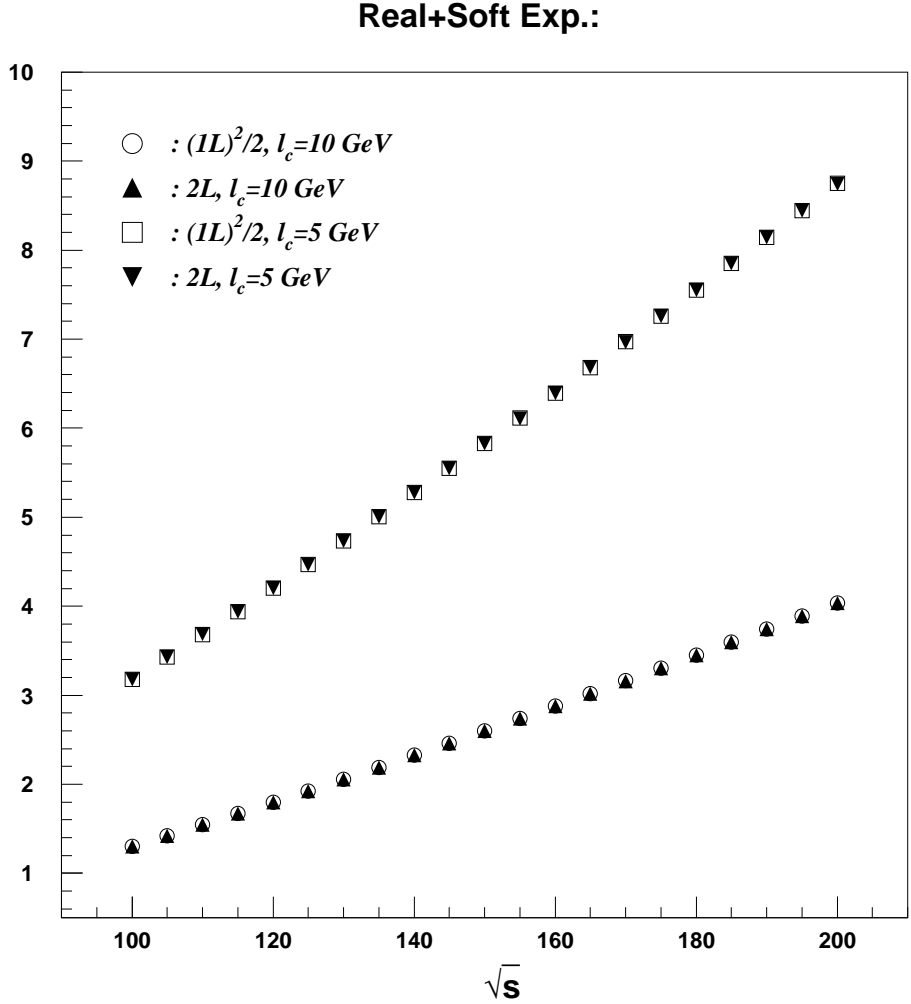,width=15cm}
\caption{The second term in the exponential obtained from the
renormalization group improved massive one-loop Sudakov form factor compared with the
explicit two-loop results given in the appendix. 
The latter depends on a different scale in each loop integration.
The quark mass is kept fixed at $m_b=4.5$~GeV and the
values for the real gluon energy cutoff $l_c$ is indicated in the figure. 
The result in Eq.~\ref{eq:RpVRG}
clearly exponentiates.}
\label{fig:cutexp}
\end{figure}
\end{center}

The DL result for the virtual one-loop Sudakov form factor can be
written in terms of integrals over $\alpha_1$ and $\beta_1$ as:
\begin{eqnarray}
{\cal F}^{DL}_{S_V} &=& - \frac{\alpha_s C_F}{2 \pi} \int^1_0
\frac{d \beta_1}{\beta_1} \int^1_0 \frac{d \alpha_1}
{\alpha_1} \Theta ( s \alpha_1 \beta_1- \lambda^2) \Theta ( \alpha_1
- \frac{m^2}{s} \beta_1) \Theta (\beta_1- \frac{m^2}{s} \alpha_1
) \nonumber \\
&=& - \frac{\alpha_s C_F}{2 \pi} \left( \frac{1}{2} \log^2 \frac{s}{m^2}
+ \log \frac{s}{m^2} \log \frac{m^2}{\lambda^2} \right) \label{eq:1lDL} \; .
\end{eqnarray}
Here $\lambda$ is a fictitious gluon mass introduced to regulate the 
infra-red divergences in the real and virtual gluon integrals.
For the inclusion of double logarithms from the real Bremsstrahlung contribution
we introduce a cutoff for the energy integration according to $\lambda \leq l_0
\leq l_c$, i.e. in terms of Sudakov variables, $\alpha_1+\beta_1
\leq \frac{l_c}{E_{\rm cm}}=\frac{2 l_c}{\sqrt{s}}$. Thus we find for the real
DL Bremsstrahlung contribution:
\begin{eqnarray}
{\cal F}^{DL}_{S_R} &=& \frac{\alpha_s C_F}{\pi} \int^1_0
\frac{d \beta_1}{\beta_1} \int^1_0 \frac{d \alpha_1}
{\alpha_1} \Theta ( s \alpha_1 \beta_1- \lambda^2) \Theta ( \alpha_1
- \frac{m^2}{s} \beta_1) \Theta (\beta_1- \frac{m^2}{s} \alpha_1
) \nonumber \\
&& \times \Theta \left( \frac{2 l_c}{\sqrt{s}} -\alpha_1-\beta_1
\right) \nonumber \\
&=& - 2{\cal F}^{DL}_{S_V} - \frac{\alpha_s C_F}{ \pi} \int^1_0
\frac{d \beta_1}{\beta_1} \int^1_0 \frac{d \alpha_1}
{\alpha_1} \Theta ( \alpha_1
- \frac{m^2}{s} \beta_1) \Theta (\beta_1- \frac{m^2}{s} \alpha_1
) \nonumber \\
&&\times \Theta \left( \alpha_1+\beta_1-\frac{2 l_c}
{\sqrt{s}} \right) \nonumber \\
&=& \frac{\alpha_s C_F}{\pi} \left( \frac{1}{2} \log^2 \frac{s}{m^2}
+ \log \frac{s}{m^2} \log \frac{m^2}{\lambda^2} - \log \frac{s}{m^2}
\log \frac{s}{4 l_c^2} \right) \label{eq:1lDLrpv}
\end{eqnarray}
assuming only $ \lambda < \frac{2l_c m}{\sqrt{s}}$. These results are of course
well known \cite{Sud,fkm,ms2} but now can be used  to insert a running
coupling into {\it each} loop integration. The DL result is sufficient for this
purpose as the RG-logarithms are sub-leading at the next order in $\alpha_s$.
At this point we only give the result for the sum  ${\cal F}^{DL}_{S_R}+
2 {\cal F}^{DL}_{S_V}$, as only this combination is relevant for a physical
cross section. We emphasize at this point that in a adopting a certain jet
definition, we need to make sure that it does not restrict the exponentiation
of the energy cut dependent piece of the soft gluon matrix elemtent. 
In the Appendix we give the result for 
the renormalization group improved ${\cal F}^{DL}_{S_V}$ form factor and in addition
the two-loop results, thus  checking the requirement that all one-loop RG-form factors
exponentiate. We find
\begin{eqnarray}
{\cal F}^{RG}_{S_R}+2 {\cal F}^{RG}_{S_V}&=& -\frac{C_F}{\pi} \left\{
\int_\frac{2l_cm^2}{(s+m^2) \sqrt{s}}^\frac{2l_c\sqrt{s}}{s+m^2} \frac{d 
\beta_1}{\beta_1} \int^1_{\frac{2l_c}{\sqrt{s}}-\beta_1}
\frac{d \alpha_1} {\alpha_1} +
\int^1_\frac{2l_c}{\sqrt{s}} \frac{d \beta_1}{\beta_1}
\int^1_{\frac{m^2}{s} \beta_1}
\frac{d \alpha_1} {\alpha_1} \right. \nonumber \\ && \left. 
- \int_\frac{2l_cm^2}{s\sqrt{s}}^\frac{m^2}{s} \frac{d \beta_1}{\beta_1}
\int^1_{\frac{s}{m^2} \beta_1}
\frac{d \alpha_1} {\alpha_1}  
\right\} \frac{\alpha_s(m^2)}{1+c \; \log \frac{ s \alpha_1 \beta_1
}{m^2}} \nonumber \\ 
&=& -\frac{\alpha_s(m^2) C_F}{\pi \;c} \left\{ - \int_\frac{2l_cm^2}{(s+m^2) \sqrt{s}}^
\frac{2l_c\sqrt{s}}{s+m^2} \frac{d \beta_1}{\beta_1} \log \frac{ 1+c \; \log
\left( \left(\frac{2l_c}{\sqrt{s}} -\beta_1 \right) \beta_1 \frac{s}{m^2}
\right)}{\left(
1+c \; \log \frac{s \beta_1}{m^2} \right)} \right. \nonumber \\
&& + \log \frac{s}{m^2} \log \frac{\alpha_s(2l_c \sqrt{s})}{
\alpha_s(s)} + \log \frac{2l_c}{\sqrt{s}} \log \frac{\alpha_s (2 l_c \sqrt{s})}
{ \alpha_s \left( \frac{2l_cm^2}{\sqrt{s}}\right)} \nonumber \\ && \left. +
\frac{1}{c} \log \frac{\alpha_s(m^2)\alpha_s(2l_c \sqrt{s})}{
\alpha_s(s) \alpha_s \left(\frac{2l_c m^2}{\sqrt{s}} \right)} \right\}   \; . 
\label{eq:RpVRG}
\end{eqnarray}
An expansion in $\alpha_s(m^2)$ gives the double logarithmic result plus
subleading terms in $\beta_0$ etc. The dependence on the cutoff $l_c$ will vanish
when the hard-gluon Bremsstrahlung contributions are added order by order. 
Fig.~\ref{fig:cutexp} compares the second term in the exponential obtained from
the renormalization group improved result in Eq.~\ref{eq:RpVRG} with the explicit
two-loop result given in Eq.~\ref{eq:RpVRG2L} of the Appendix.\footnote{The 
two-loop numerical result is
obtained by Monte Carlo integration with $10^6$ evaluations for each of the
fifty iterations.} We emphasize that the two loop result depends on two 
``running'' scales , ${\bf l_{1_\perp}}$ and ${\bf l_{2_\perp}}$. 
The figure clearly shows that our one-loop result exponentiates as
expected.

Since in this
work we are  only able  to include the hard {\it one-gluon} matrix elements
(the exact NNLO corrections are presently unknown),
the higher-order terms are at this point of a more academic interest. 
Our two jet definition (see below) restricts the higher-order hard gluon radiation phase
space sufficiently so that it is reasonable to neglect more than one gluon emission.
As mentioned above, Fig. \ref{fig:cutexp} shows, 
using explicit RG-improved two loop DL-results
in the Appendix, that the result in Eq.~\ref{eq:RpVRG} exponentiates 
as expected. We do have to consider the subleading terms already
entering at the one-loop level as the hard Bremsstrahlung contribution will
also contain subleading terms in $l_c$. The complete result for the renormalization
group improved massive Sudakov form factor is thus given by:
\begin{eqnarray}
\widetilde{{\cal F}}^{RG}_{S_R}+2 \widetilde{{\cal F}}^{RG}_{S_V}&=& 
\frac{\alpha_s(m^2) C_F}{\pi} \left\{ \frac{1}{c}
\int_\frac{2l_cm^2}{(s+m^2) \sqrt{s}}^
\frac{2l_c\sqrt{s}}{s+m^2} \frac{d \beta_1}{\beta_1} \log \frac{ 1+c \; \log
\left( \left(\frac{2l_c}{\sqrt{s}} -\beta_1 \right) \beta_1 \frac{s}{m^2}
\right)}{\left(
1+c \; \log \frac{s \beta_1}{m^2} \right)} \right. \nonumber \\
&& - \frac{1}{c} \log \frac{s}{m^2} \log \frac{\alpha_s(2l_c \sqrt{s})}{
\alpha_s(s)} - \frac{1}{c}\log \frac{2l_c}{\sqrt{s}} \log \frac{\alpha_s (2 l_c \sqrt{s})}
{ \alpha_s \left( \frac{2l_cm^2}{\sqrt{s}}\right)} \nonumber \\ && \left. -
\frac{1}{c^2} \log \frac{\alpha_s(m^2)\alpha_s(2l_c \sqrt{s})}{
\alpha_s(s) \alpha_s \left(\frac{2l_c m^2}{\sqrt{s}} \right)} 
+ \frac{1}{2} \log \frac{s}{m^2} + \log \frac{s}{4 l_c^2} -1 + \frac{\pi^2}{3} \right\} 
\; ,
\label{eq:RpVRGex}
\end{eqnarray}
assuming only $\frac{m^2}{s} \ll 1$. Expanding in $\alpha_s(m^2)$ gives the
DL-Sudakov form factor in Eq.~\ref{eq:vps} together with  subleading terms proportional
to $\beta_0$ and subsubleading terms proportional to $\beta_1$. We emphasize
that the two-loop running coupling is included in Eq.~\ref{eq:RpVRGex} to
all orders and that all collinear divergences are avoided due to the fact that
we keep all non-homogeneous fermion mass terms.
We next turn to the virtual hard DL
corrections and investigate the RG effects for these
contributions.

\subsection{The Non-Sudakov RG-Form Factor}

\begin{center}
\begin{figure}
\centering
\epsfig{file=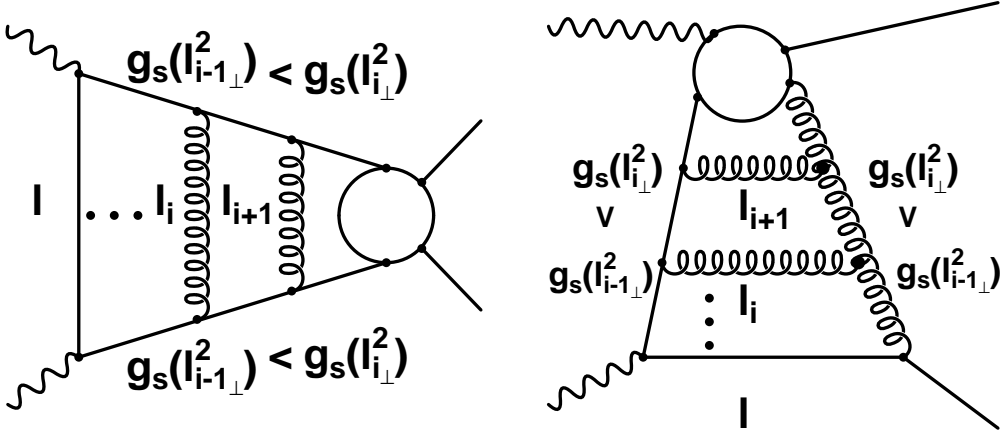,width=15cm}
\caption{The schematic Feynman diagrams leading to the hard (non-Sudakov)
double logarithms in the process $\gamma \gamma (J_z=0) \longrightarrow q 
\overline{q}$ with $i+1$ gluon insertions. 
The blobs denote a hard momentum going through the omitted propagator
in the DL-phase space.
Crossed diagrams lead to a different
ordering of the Sudakov variables and are correctly accounted for by a factor
of $(i+1)!$ at each order. The scale of the coupling $\alpha_s=\frac{g_s^2}{4
\pi}$ is indicated at the vertices and included explicitly in this work. The topology on 
the left-hand diagram is Abelian like,  and the one on the right is
non-Abelian beyond one loop.}
\label{fig:RG}
\end{figure}
\end{center}

In the absence of exact two-loop QCD corrections to the process $\gamma \gamma
(J_z=0) \longrightarrow q \overline{q}$ it might seem unclear how to perform
a renormalization group improvement for the hard non-Sudakov DL's. We find,
however, 
following similar arguments as in the case of the massive Sudakov DL's of the
previous section,  that the same effective scale ${\bf l^2_\perp}$ is also
valid for
the novel non-Sudakov logarithms which were summed to give  the confluent 
hypergeometric function $_2F_2$ in Ref.~\cite{ms}. One way to see this is
to note that  on a formal level, all insertions of gluons into the
hard topology shown in Fig.~\ref{fig:RG} have the same structure as in the
Sudakov case. Only the last fermion loop integration separates the two cases
by effectively regularizing soft divergences with the quark mass. The strong
coupling receives no renormalization from this last integration so that
the scales of the couplings at each order are determined by the same
renormalization group arguments as for the Sudakov case. In this context it is
worth pointing out that our analysis here also thus removes the largest 
uncertainty in the exact one-loop calculation of Ref.~\cite{jt}, as the
scale of $\alpha_s$ in that work was unspecified at this order.

\begin{center}
\begin{figure}
\centering
\epsfig{file=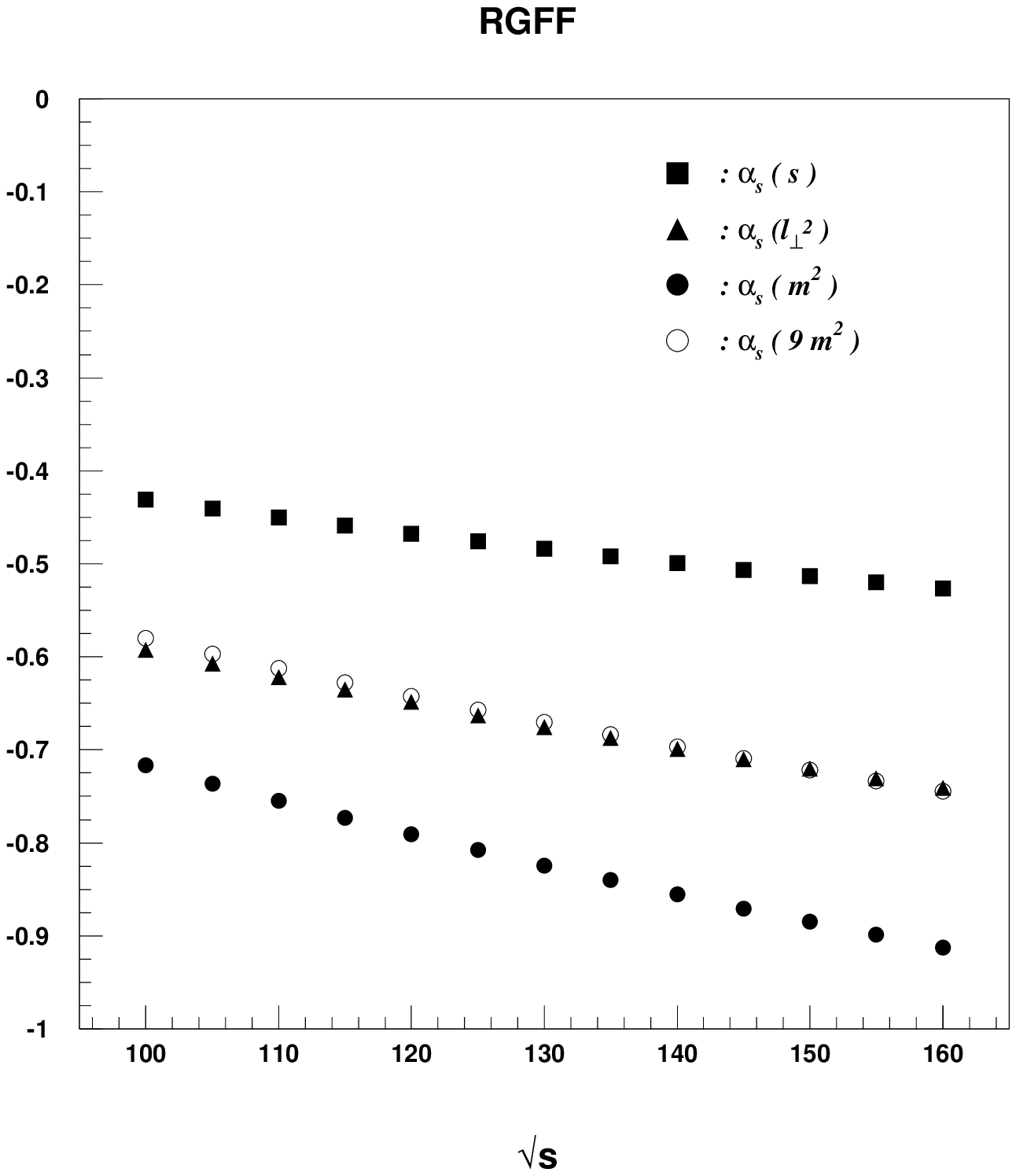,height=15cm}
\caption{The effect of incorporating a running coupling constant at each loop
integration according to Eq.~\ref{eq:rgres}. The values of the form factor
in the DL approximation of Ref.~\cite{ms} are also shown with their indicated
scale of where $\alpha_s$ was evaluated. Choosing an effective scale
of roughly $9 m^2$, one obtains results very close to the exact renormalization
group improved values.}
\label{fig:rgff}
\end{figure}
\end{center}

\begin{center}
\begin{figure}
\centering
\epsfig{file=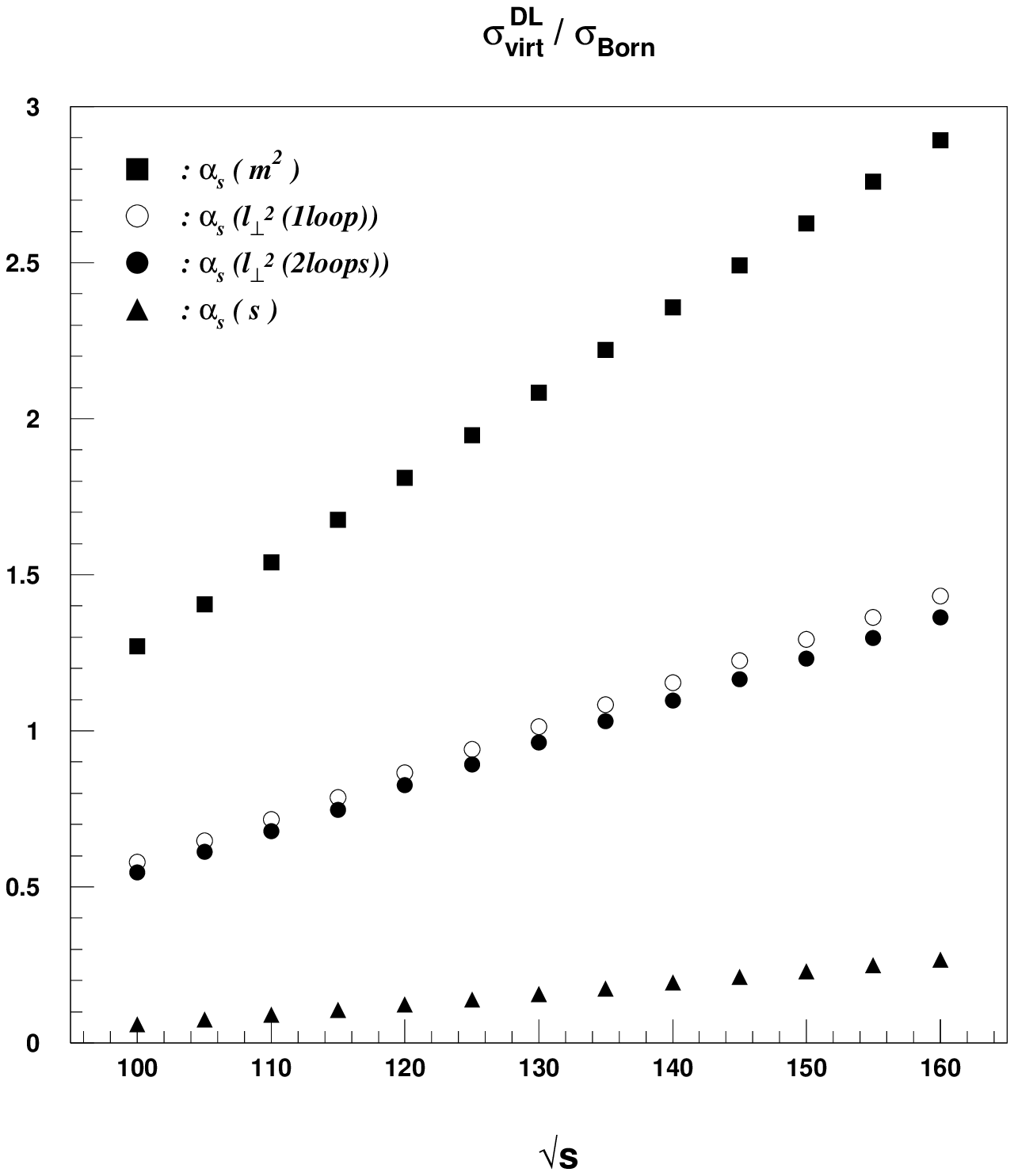,height=15cm}
\caption{The effect of incorporating a running coupling constant at each loop
integration (circles) according to Eq.~\ref{eq:srg}
(suppressing the Sudakov term). The one- and two-loop
running coupling solutions are within $8\%$ of each other.
Also shown are upper and lower
limits according to the indicated values of $\alpha_s$ in the DL approximation.}
\label{fig:srg}
\end{figure}
\end{center}

From the exact next-to-leading order result for the running coupling in 
Eq.~\ref{eq:rc}
it is clear that a formulation in terms of ${\bf l^2_\perp}$ of the series leading
to the non-Sudakov double logarithms is more adaptable to a renormalization
group improvement. In Ref.~\cite{ms} these contributions were derived by
first integrating over all $l^2_\perp$ with integrals over $\{\alpha, \beta \}$
pairs remaining. For the topology depicted on the left in Fig.~\ref{fig:RG} 
we find that it can be reformulated as follows:
\begin{eqnarray}
{\cal F} \; _2F_2 (1,1;2,\frac{3}{2};\frac{1}{2} {\cal F}) \!\!\! &=& \!\!\!\!
\sum^\infty_{n=1}
\int^1_{\frac{m^2}{s}} \frac{d \alpha}{\alpha} \int^1_{\frac{m^2}{s \alpha}}
\frac{d \beta}{\beta} \prod^{n-2}_{i=0} \Gamma(n) \!\! \int^1_{\alpha_i} \frac{d \alpha_{i+1}}
{\alpha_{i+1}} \int^1_{\beta_i} \frac{d \beta_{i+1}}{\beta_{i+1}} 
\left( \frac{- \alpha_s C_F}{2 \pi} \right)^n \nonumber \\
&=& \!\!\!\! \sum^\infty_{n=1}
\int^s_{m^2} \frac{d {\bf l^2_\perp}}{{\bf l^2_\perp}} \int^1_{\frac{{\bf l^2_\perp}}{s}}
\frac{d \alpha}{\alpha} \prod^{n-2}_{i=0} \Gamma(n) \!\! \int^{{\bf l^2_{i_\perp}}}_{m^2} 
\! \frac{d {\bf l^2_{{i+1}_\perp}}}
{{\bf l^2_{{i+1}_\perp}}} \int^1_{\alpha_i} \frac{d \alpha_{i+1}}{\alpha_{i+1}} 
\left( \frac{-\alpha_s C_F}{2 \pi} \right)^n \label{eq:lpdl}
\end{eqnarray}
where ${\cal F}= - \frac{\alpha_s C_F}{4 \pi} \log^2 \frac{s}{m^2}$ denotes the hard
one-loop form factor of Ref.~\cite{ms}. In this result  
the product  is set to one for $n=1$, and contains nested integrals for
$n \ge 2$ with ${\bf l^2_{0_\perp}} \equiv {\bf l^2_\perp}$
and $\alpha_0 \equiv \alpha$. From this expression it is clear that
an incorporation of the running coupling in Eq.~\ref{eq:rc} will not contain 
any Landau pole singularity \cite{cm} as $m^2 \leq {\bf l^2_{i_\perp}} \leq {\bf l^2_\perp}$.
We now include the running of $\alpha_s$
according to Eq.~\ref{eq:rc} as follows. For each gluon insertion we have
\begin{eqnarray}
\int^{{\bf l^2_{i_\perp}}}_{m^2} \frac{d {\bf l^2_{{i+1}_\perp}}}
{{\bf l^2_{{i+1}_\perp}}} \int^1_{\alpha_i} \frac{d \alpha_{i+1}}{\alpha_{i+1}} \frac{\alpha_s
(m^2)}{1+ c
\; \log \frac{{\bf l^2_{{i+1}_\perp}}}{m^2}}
&=&- \frac{\alpha_s(m^2)}{c} 
\log \alpha_i \log (1+ c \;
\log 
\frac{{\bf l^2_{i_\perp}}}{m^2}) \nonumber \\
&=&-\frac{\alpha_s(m^2)}{c}
\log \alpha_i \log \frac{\alpha(m^2)}{\alpha({\bf l^2_{i_\perp}})} \; .
\label{eq:gi}
\end{eqnarray}
With
\begin{equation}
\frac{ d {\bf l^2_{i_\perp}}}{{\bf l^2_{i_\perp}}} =  
- \frac{\alpha_s(m^2)}{c} 
\frac{ d \log ( \alpha_s
({\bf l^2_{i_\perp}}))}{\alpha_s ( {\bf l^2_{i_\perp}} )}
\end{equation}
we find
\begin{equation}
\int^{{\bf l^2_{{i-1}_\perp}}}_{m^2} \frac{ d {\bf l^2_{i_\perp}}}{{\bf l^2_{i_\perp}}} \alpha_s (
{\bf l^2_{i_\perp}}) \left( 
- \frac{\alpha_s(m^2)}{c}
\log \frac{\alpha(m^2)}
{\alpha({\bf l^2_{i_\perp}})} \right) = 
\frac{\alpha^2_s(m^2)}{2 \; c^2}
\log^2 \frac{\alpha(m^2)}{\alpha({\bf l^2_{{i-1}_\perp}})} \; .
\end{equation}
It is clear from this derivation that for $i$-gluon iterations we have
\begin{equation}
\frac{(-1)^{i+1}}{i!} \left( 
\frac{\alpha_s(m^2)}{ c}
\right)^i \frac{\log^i \alpha}{\alpha}
\log^i \frac{\alpha_s ( {\bf l^2_\perp} )}{\alpha_s ( m^2 )} \; .
\end{equation}
Thus we finally arrive at the complete renormalization group improved result
for the hard non-Sudakov form factor corresponding to the left 
(``Abelian'') topology in Fig.~\ref{fig:RG}:
\begin{eqnarray}
{\cal F}_h^{RG}&=&\sum^\infty_{i=0} \int^s_{m^2} \frac{ d {\bf l^2_\perp}}{{\bf l^2_\perp}}
\int^1_{\frac{{\bf l^2_\perp}}{s}} \frac{d \alpha}{\alpha} \left(\frac{-C_F
}{2 \pi} \right)^{i+1} \left( 
\frac{\alpha_s(m^2)}{ c} \right)^i
\frac{\alpha_s({\bf l^2_\perp})}{i!}
\frac{\log^i \alpha}{\alpha}  \log^i 
\frac{\alpha_s ( {\bf l^2_\perp} )}
{\alpha_s ( m^2 )} \nonumber \\
&=&\sum^\infty_{i=0} \int^s_{m^2} \frac{ d {\bf l^2_\perp}}{{\bf l^2_\perp}}
\left(\frac{C_F
}{2 \pi} \right)^{i+1} \left(
\frac{- \alpha_s(m^2)}{ c} \right)^i
\frac{\alpha_s({\bf l^2_\perp})}{(i+1)!}
\log^{i+1} \frac{{\bf l^2_\perp}}{s} \log^i 
\frac{\alpha_s ( {\bf l^2_\perp} )}
{\alpha_s ( m^2 )} \; . \label{eq:rgres}
\end{eqnarray}
The effect of the renormalization group improved hard form factor is shown in Fig.
\ref{fig:rgff} for the case of the $b$-quark. The effective scale of the 
coupling in the DL approximation\footnote{By this we mean the scale which when 
used for $\alpha_s$ in the non-RG-improved DL result reproduces numerically
the RG-improved result derived here.} is roughly $9 m_b^2$.
For the series obtained in the non-Abelian topology on the right of 
Fig.~\ref{fig:RG} only the
color factor changes,  as was shown in Ref.~\cite{ms}. The correct
substitution at the $(i+1)$-loop level is
\begin{equation}
C^{i+1}_F \longrightarrow C_F \left( \frac{C_A}{2} \right)^i
\end{equation}
with an additional  factor of $2$ as this topology occurs twice in the process
$\gamma \gamma \longrightarrow q \overline{q}$. In summary, the complete
virtual renormalization group improved hard form factor is  given by
\begin{eqnarray}
\widetilde{{\cal F}}_h^{RG}
&=&\sum^\infty_{i=0} \int^s_{m^2} \frac{ d {\bf l^2_\perp}}{{\bf l^2_\perp}}
\left(\frac{C_F
}{2 \pi} \right)^{i+1} \left(
\frac{\alpha_s(m^2)}{ c}\right)^i
\frac{\alpha_s({\bf l^2_\perp})}{(i+1)!}
\log^{i+1} \frac{{\bf l^2_\perp}}{s} \log^i 
\frac{\alpha_s(m^2)}{\alpha_s ( {\bf l^2_\perp} )}
+  \nonumber \\
&&2 \sum^\infty_{i=0} \int^s_{m^2} \frac{ d {\bf l^2_\perp}}{{\bf l^2_\perp}}
\frac{C_F C_A^i
}{2^{2i+1} \pi^{i+1}} \left(
\frac{\alpha_s(m^2)}{c} \right)^i
\frac{\alpha_s({\bf l^2_\perp})}{(i+1)!}
\log^{i+1} \frac{{\bf l^2_\perp}}{s} \log^i 
\frac{\alpha_s(m^2)}{\alpha_s ( {\bf l^2_\perp} )} \; ,
\label{eq:rgff}
\end{eqnarray}
and thus
\begin{equation}
\frac{ \sigma^{DL}_{RG}}{\sigma_{\rm Born}} = \left\{ 1 + \widetilde{{\cal F}}_h^{RG}
\right\}^2 \exp \left( \widetilde{{\cal F}}^{RG}_{S_R} + 2 \widetilde{{\cal F}}^{RG}_{S_V}
\right) \label{eq:srg}
\end{equation}
where the RG-improved massive Sudakov form factor is given in Eq.~\ref{eq:RpVRGex}.
Fig.~\ref{fig:srg} shows the size of the effect at the cross-section level
for $m=m_b=4.5$~GeV. We take  $\alpha_s(m_b^2)=0.2$ and use Eq.~\ref{eq:rc}
with $n_F=5$ to arrive at $\alpha_s(s)$. The effect is significant and lies 
almost in the middle of the `theoretical' upper and lower limits 
for the non-Sudakov form factor (suppressing the Sudakov term for now)
as indicated
in the figure.
In the following section we will now investigate the effect on the full cross
section, including the Sudakov form factor as well as the exact one loop radiative
corrections.

\section{Numerical Results} \label{sec:nr}

In Ref.~\cite{ms2} we presented numerical predictions for an (infra-red safe)
two-jet $b \bar b$ cross section in $e^+e^-$ collisions in the energy
range $\sqrt{s} = 100  - 200$~GeV. We used a modified 
Sterman-Weinberg cone definition. Thus, at leading order
(i.e. $\gamma\gamma \to b \overline{b}$) all events obviously
satisfy the two--$b$--jet
requirement.\footnote{We apply an angular cut of
$\vert{\cos\theta_{b,\overline{b}}}\vert
< 0.7$ to ensure that both jets lie in the central region.} This defines
our `leading order' (LO) cross section. 
At next--to--leading order (NLO) we can have virtual or real gluon emission.
For the latter, an event is defined as two--$b$--jet like if the emitted
gluon 
\begin{eqnarray*}
\mbox{{\it either}}&& \mbox{I.\quad has energy less than $\epsilon \sqrt{s}$, with  
$\epsilon \ll 1$}, \\
\mbox{{\it or}} && \mbox{II.\quad is within an angle 
$2 \delta$ of the $b$ or $\overline{b}$, again
with $\delta \ll 1$}.
\end{eqnarray*}
 we further subdivided 
region I according to whether
the gluon energy is greater or less than the infrared cutoff $l_c$
 ($< \epsilon$). Adding the virtual gluon corrections to this latter
(soft)  contribution, to give $\sigma_{\rm SV}$, and calling the remaining
hard gluon contribution $\sigma_{\rm H}$, we have 
\begin{equation}
\sigma_{\rm 2j} = \sigma_{\rm SV}(l_c) + 
\sigma_{\rm H}(l_c,\epsilon,\delta) \; .
\label{eq:2jet}
\end{equation}
In Ref.~\cite{ms2} we evaluated each part of this cross section exactly
to ${\cal O}(\alpha_s)$ and in addition we included the resummed
hard non-Sudakov form factor in $\sigma_{\rm SV}$. This was necessary
to yield a positive cross section.

We can now update these predictions using the RG-improved expressions
for the resummed form factors. Thus 
\begin{equation}
\sigma_{\rm SV } \longrightarrow \sigma^{DL}_{RG} + 
\tilde\sigma_{\rm SV }\; ,
\end{equation}
where $\sigma^{DL}_{RG}$ is given in Eq.~\ref{eq:srg} and 
$\tilde\sigma_{\rm SV }  $ is the exact one-loop result minus the one loop 
leading-logarithm pieces which are resummed in $ \sigma^{DL}_{RG}$, i.e.
\begin{eqnarray}
\tilde\sigma_{\rm SV }  &=& \sigma_{\rm SV, NLO} - 
\sigma_{\rm LO} \left[ -6 \cF  +  \frac{ \alpha_s C_F}{\pi} 
\left( \log \frac{s}{m_q^2} \left(
\frac{1}{2} - \log \frac{s}{4 l_c^2} \right) + \log \frac{s}{4 l_c^2} -1 +
\frac{\pi^2}{3} \right) \right] \; . \nonumber \\
&&
\end{eqnarray}
By adding the second (Sudakov) piece in the square brackets we remove
(at least up to terms $\cO(l_c^2/s) \ll 1$) the dependence on the gluon
energy cutoff $l_c$. 
Note also that the complete expression for the two-jet cross section 
(with the remaining $l_c$ dependence displayed)
\begin{equation}
\sigma_{\rm 2j} = \sigma^{DL}_{RG}(l_c) + \tilde\sigma_{\rm SV} + 
\sigma_{\rm H}(l_c,\epsilon,\delta) \; .
\label{eq:2jetbis}
\end{equation}
contains  a mixture of exact ${\cal O}(\alpha_s)$ and resummed pieces.
For the former, we use $m_b^2$ as the scale for $\alpha_s$.\footnote{We chose
the QCD scale parameter $\Lambda$ such that $\alpha_s(m_b^2)=0.2$
for $m_b = 4.5$~GeV, at both leading and next-to-leading
order.} We believe
that this is a more reasonable physical 
choice than the fixed value $\alpha_s = 0.11$ used for illustrative
purposes in \cite{ms2}. The resummed contributions are based on the 
scale choice ${\bf l^2_\perp}$ in the loops, as already discussed.

Before computing  and combining the various components of the two-jet cross section
in Eq.~\ref{eq:2jetbis} we must address the issue of the dependence on the unphysical
infra-red parameter $l_c$. If we were to expand out the resummed RG-improved form factor
$\sigma^{DL}_{RG}(l_c)$ in powers of $\alpha_s(m_q^2)$, and retain only
the $\cO(\alpha_S)$ term, we would find that the $l_c$ dependence exactly canceled
that of $\sigma_{\rm H}(l_c,\epsilon,\delta)$.\footnote{This was shown explicitly
in Ref.~\cite{ms2}, see for example Fig.~3 therein.} However in the full resummed expression,
there is nothing to cancel the explicit $l_c$ dependence at higher-orders. The canceling
terms would come from the as yet unknown higher-order contributions to $\sigma_{\rm H}$.
Faced with this dilemma, we have several choices. We could, as in \cite{ms2}, neglect
the higher-order terms in the 
Sudakov form factor altogether, and include only the non-Sudakov form factor
which is of course independent of $l_c$. Furthermore, we showed in \cite{ms2}
that with the choice $\epsilon=\cO(0.1)$, the combined 
contribution of virtual gluons and real gluons with $E_g < \epsilon \sqrt{s}$
to $\sigma_{\rm 2j}$ was dominated by the non-Sudakov ``$6 \cF$" part. This 
suggests that the most reasonable procedure for the resummed
cross section is to take $l_c \sim \epsilon \sqrt{s}$. We stress that this 
is an {\it approximation}, since it corresponds to making an assumption
about the contribution of real multi-gluon emission with energies 
$ < \cO(\epsilon\sqrt{s})$.

\begin{center}
\begin{figure}
\centering
\epsfig{file=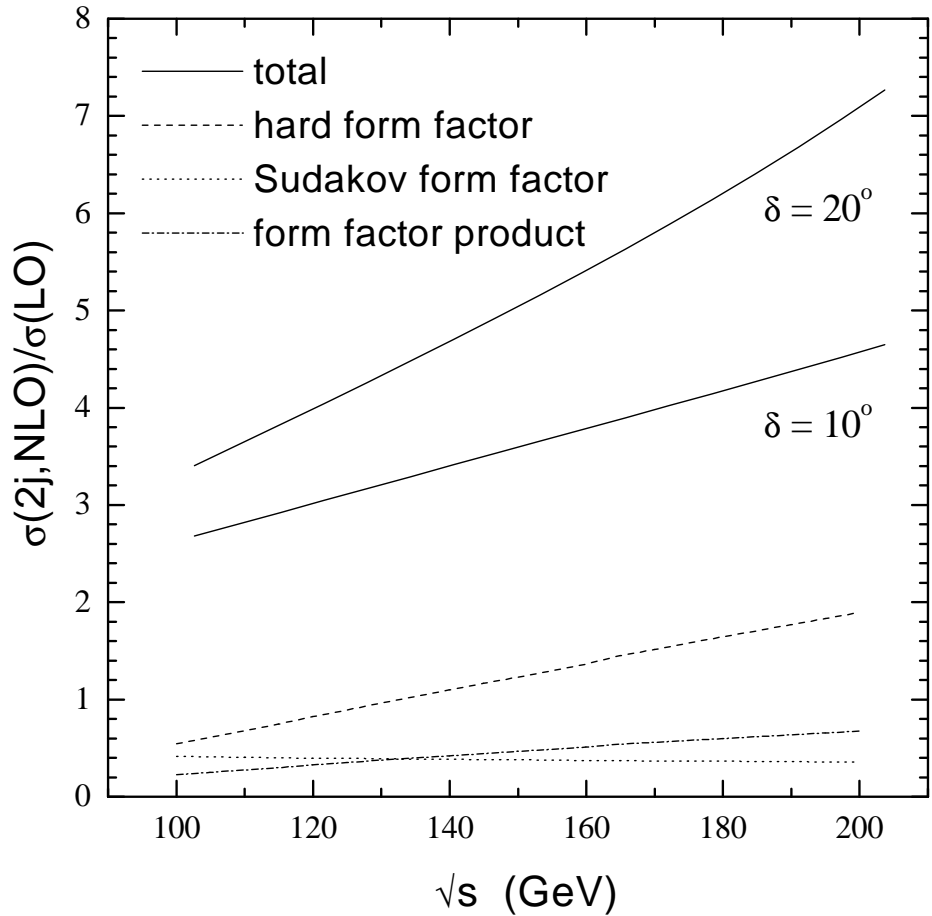,width=15cm}
\caption{The  total two--$b$--jet
cross section (i.e. exact next--to--leading order contribution
plus resummed Sudakov and non-Sudakov  form factors) normalized to leading order, 
for the  Sterman-Weinberg jet parameter choices $(\epsilon = 0.1,\delta=
10^\circ$ and $20^\circ)$.
Also shown  is the separate non-Sudakov  (dashed line) and
Sudakov (dash-dotted line) form factor contributions, and their product
(long-dashed line).}
\label{fig:total}
\end{figure}
\end{center}

As our `best guess' RG-improved, resummed two-jet cross section, therefore, we have
\begin{equation}
\sigma_{\rm 2j} = \sigma^{DL}_{RG}(\epsilon\sqrt{s}) + \tilde\sigma_{\rm SV} + 
\sigma_{\rm H}(\epsilon\sqrt{s},\epsilon,\delta) \; .
\label{eq:2jetbisbis}
\end{equation}
Figure~\ref{fig:total} shows the $\sqrt{s}$ dependence of this cross section,
normalized to the leading-order (Born) cross section. Evidently, the effect
of the renormalization group improved results is significant when compared
to the results in Ref. \cite{ms2} (where the scale of $\alpha_s$ was a free
parameter). The Sudakov form factor reduces the all orders hard RG-form factor
by over a factor of two, although it is still roughly a factor of three larger
than if $\alpha_s(s)$ had been chosen in the hard DL-form factor. The hard
Bremsstrahlung contribution is strongly dependent on the available phase space
as the mass suppression is removed for these terms. The size of the
Bremsstrahlung corrections is thus enhanced substantially through the large
effective scale of the RG-Sudakov form factor.
In total, we find that the RG effects lead to an {\it enhancement} of the
background of roughly a factor of two compared with the scale chosen in
Ref. \cite{ms2}. 

We note, however, that a similar effect for the Sudakov form factor is also
present in the Signal $H \longrightarrow b \overline{b}$, so that the
RG-effects in the non-Sudakov form factor-and in the misidentified
$c\overline{c}$ background both Sudakov {\it and} non-Sudakov-
will be decisive in a precise
determination of the Signal/BG ratio.

\section{Conclusions} \label{sec:con}

In this paper we have shown how to incorporate the exact one and two loop running
coupling into the massive Sudakov as well as the hard non-Sudakov double 
logarithmic form factors found in 
$\gamma \gamma (J_z=0)\longrightarrow q \overline{q}$. 
This process is very important at a 
future PLC as it contains the largest background to intermediate mass Higgs
production. 
The RG-improvement is achieved without encountering
the ominous Landau-pole singularities in the hard form factor although the all orders leading and
next to leading RG-solution is incorporated. The reason for this fortuitous fact
is that the novel non-Sudakov form factors are ``regulated'' 
by the fermion mass,
yielding an effective low energy cutoff. 

The size of the effect is significant and is found to stay within
the theoretical upper and lower limits for both the form factor as well as
the cross section contribution. The background is enhanced significantly
compared to Ref.~\cite{ms2} by the renormalization
group effects as the effective scale is found to be roughly $9 m^2$ for the 
hard form factor. The effective scale for the RG-Sudakov form factor is 
cutoff dependent.

In this context it is worth pointing out a major difference to 
the $ \gamma \gamma
\longrightarrow H$ signal contribution from a b-quark loop. 
There a function similar to the Abelian form factor
of Eq.~\ref{eq:lpdl} occurs, as was shown in Ref.~\cite{ky}. 
There is, however, a big difference with respect to the renormalization
of the strong coupling, as in the case of $ \gamma \gamma \longrightarrow H$
$\alpha_s$ is only renormalized at the three-loop level. The effect of the
running coupling in that process is thus much smaller, but perhaps still
significant in the case of a large $tan \beta$ enhancement scenario \cite{m}.

The effect of a running mass parameter is subleading as it occurs only as
the arguments of logarithms. The mass term in the factorized Born amplitude
is fixed by the on-shell renormalization scheme employed in the exact one
loop calculation of Refs.~\cite{jt,jt2}.

A full Monte Carlo study of both signal and background effects including
investigations of optimal experimental cuts  and $b$-tagging efficiencies 
is work in progress \cite{kms}.

\noindent{\bf Acknowledgements}\\ 
We would like to thank M.~W\"usthoff for discussions.
This work was supported in part by the EU Fourth Framework Programme `Training and Mobility of 
Researchers', Network `Quantum Chromodynamics and the Deep Structure of Elementary Particles', 
contract FMRX-CT98-0194 (DG 12 - MIHT).  
 
\section{Appendix}

In this appendix we give explicit results for the gluon mass dependent renormalization 
group improved massive virtual Sudakov form factor. We also give two loop results
for both the virtual as well as the sum of the virtual plus soft real contribution
and show numerically that both results exponentiate.

\subsection{The massive virtual RG-improved Sudakov form factor}
 
\begin{center}
\begin{figure}
\centering
\epsfig{file=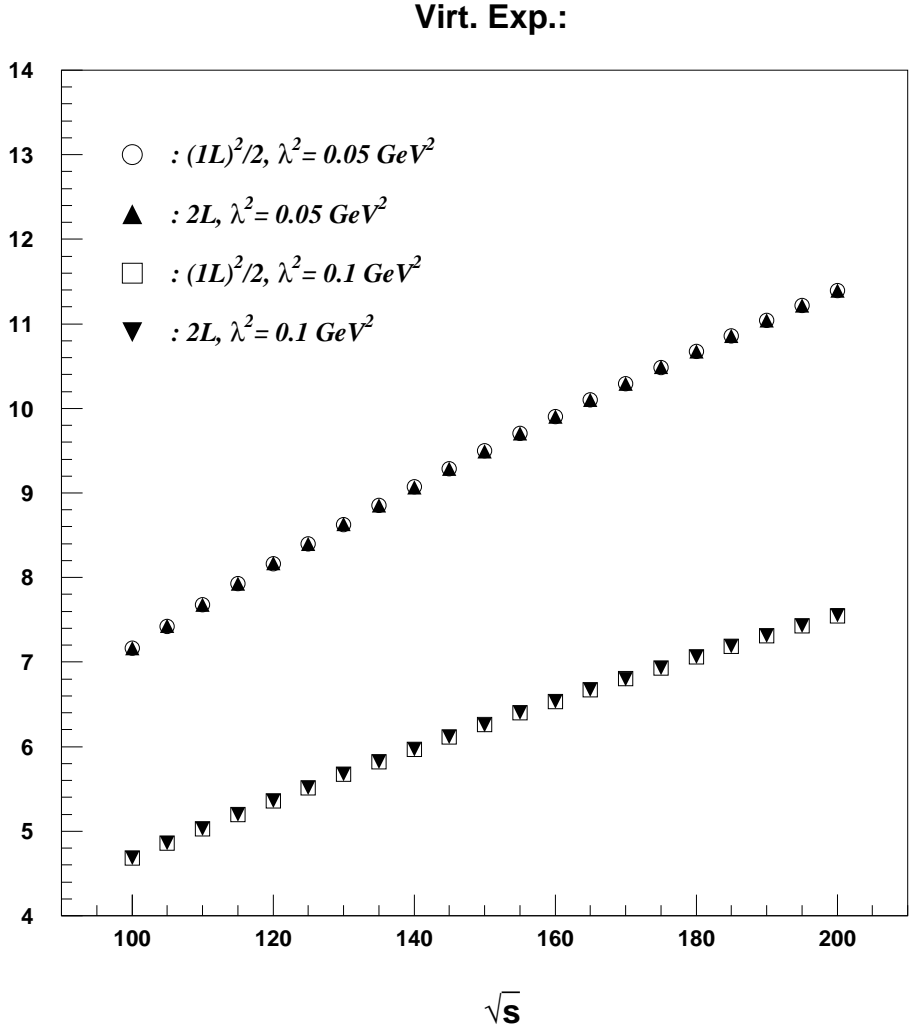,width=15cm}
\caption{A comparison of the second term in the exponential obtained from the
renormalization group improved massive one-loop Sudakov form factor compared with the
explicit two loop results. The quark mass is kept fixed at $m_b=4.5$ GeV and the
value for the gluon mass $\lambda$ is indicated in the figure. The result in Eq. \ref{eq:SvRG}
clearly exponentiates.}
\label{fig:lamexp}
\end{figure}
\end{center}

We begin with the result for ${\cal F}^{RG}_{S_V}$:
\begin{eqnarray}
{\cal F}^{RG}_{S_V} &=& - \frac{C_F}{2 \pi} \int^1_0
\frac{d \beta_1}{\beta_1} \int^1_0 \frac{d \alpha_1}
{\alpha_1} \Theta ( s \alpha_1 \beta_1- \lambda^2) \Theta ( \alpha_1
- \frac{m^2}{s} \beta_1) \Theta (\beta_1- \frac{m^2}{s} \alpha_1
) \nonumber \\
&& \times \frac{\alpha_s(m^2)}{1+c \; \log \frac{s \alpha_1 \beta_1}{m^2}}
\nonumber \\ &=&
-\frac{C_F}{2 \pi} \left\{ \int^\frac{\lambda}{m}_\frac{\lambda^2}{s} 
\frac{d \beta_1}{\beta_1} \int^1_\frac{\lambda^2}{
s \beta_1} \frac{d \alpha_1}{\alpha_1} +
\int^1_\frac{\lambda}{m} \frac{d \beta_1}{\beta_1} \int^1_
{\frac{m^2}{s}\beta_1} \frac{d \alpha_1}{\alpha_1} \right. \nonumber
\\ && \left. - \int^\frac{\lambda m}{s}_\frac{\lambda^2}{s} \frac{d 
\beta_1}{\beta_1}
\int^1_\frac{\lambda^2}{s\beta_1} \frac{d \alpha_1}{\alpha_1}
- \int^\frac{m^2}{s}_\frac{\lambda m}{s} \frac{d \beta_1}{\beta_1}
\int^1_{\frac{s}{m^2}\beta_1} \frac{d \alpha_1}{\alpha_1} \right\} 
\frac{\alpha_s(m^2)}{1+c \;\log \frac{s \alpha_1 \beta_1}{m^2}} \nonumber \\
&=& -\frac{\alpha_s(m^2) C_F}{2 \pi } \left\{ \frac{1}{c} \log \frac{s}{m^2}
\left( \log \frac{\alpha_s(\lambda^2)}{\alpha_s
(s)} - 1 \right) + \frac{1}{c^2}
\log \frac{\alpha_s(m^2)}{\alpha_s(s)} \right\}
\label{eq:SvRG}
\end{eqnarray}
The $\lambda$-dependent terms cancel out of any physical cross section when the
real Bremsstrahlung contributions are added. In order to demonstrate that the
result in Eq. \ref{eq:SvRG} exponentiates, we calculate the explicit two loop
renormalization group improved massive virtual Sudakov corrections,
now containing a different ``running scale'' in each loop, and find:
\begin{eqnarray}
{\cal F}^{{RG}_{2L}}_{S_V} &=& \frac{C^2_F}{4 \pi^2} \int^1_0
\frac{d \beta_1}{\beta_1} \int^1_0 \frac{d \alpha_1}
{\alpha_1}  \int^1_0 \frac{d \beta_2}{\beta_2}
\int^1_0 \frac{d \alpha_2}{\alpha_2} 
\frac{\alpha_s(m^2)}{1+c \;\log \frac{s \alpha_1 \beta_1}{m^2}}
\frac{\alpha_s(m^2)}{1+c \; \log \frac{s \alpha_2 \beta_2}{m^2}}
\nonumber \\ 
&\times& \left\{ \Theta ( s \alpha_1 \beta_1- \lambda^2) \Theta ( \alpha_1
- \frac{m^2}{s} \beta_1) \Theta (\beta_1- \frac{m^2}{s} \alpha_1
) \Theta ( s \alpha_2 \beta_2 - \lambda^2) \right. \nonumber \\
&&\;\;\, \Theta ( \alpha_2- \frac{m^2}{s} \beta_2) \Theta 
(\beta_2- \frac{m^2}{s} \alpha_2) \Theta (\alpha_2 - \alpha_1) 
\Theta (\beta_2-\beta_1) \;\; + \nonumber \\ && \;\;\,
\Theta ( s \alpha_1 \beta_1- \lambda^2) \Theta ( \alpha_1
- \frac{m^2}{s} \beta_1) \Theta (\beta_1- \frac{m^2}{s} \alpha_2
) \Theta ( s \alpha_2 \beta_2- \lambda^2) \nonumber \\
&& \;\;\, \left. \Theta ( \alpha_2- \frac{m^2}{s} \beta_1) \Theta
(\beta_2- \frac{m^2}{s} \alpha_2) \Theta (\alpha_2 - \alpha_1) 
\Theta (\beta_1-\beta_2) \right\} \label{eq:V2LRG}
\end{eqnarray}
These integrals cannot be solved analytically anymore, however, numerically with
the Monte Carlo generator Vegas \cite{l} it is straightforward. Fig. \ref{fig:lamexp}
displays the numerical results of Eq. \ref{eq:V2LRG} in comparison with 
half of the square of the RG-improved one loop results of Eq. \ref{eq:SvRG}.
The agreement is well within the statistical uncertainty and thus demonstrates
that also the new running coupling contributions to the massive virtual Sudakov 
form factor exponentiates as expected. For completeness we also give the 
subleading terms of the pure one loop form factor which is of course also
important for phenomenological applications. The complete result is thus given
by:
\begin{eqnarray}
\widetilde{{\cal F}}^{RG}_{S_V}
&=& -\frac{\alpha_s(m^2) C_F}{2 \pi } \left\{ \frac{1}{c} \log \frac{s}{m^2}
\left( \log \frac{\alpha_s(\lambda^2)}{\alpha_s
(s)} - 1 \right) + \frac{1}{c^2}
\log \frac{\alpha_s(m^2)}{\alpha_s(s)} \right. \nonumber \\
&& \left.
 -\frac{1}{2} \log \frac{s}{m^2} - \log \frac{ m^2}{
\lambda^2} + 1 - \frac{2 \pi^2}{3} \right\} \label{eq:SvRGex}
\end{eqnarray}

\subsection{Two loop results for the $\lambda$-independent RG-Sudakov form factor}

We now give the two loop results for the physical combination 
${\cal F}^{RG}_{S_R}+2{\cal F}^{RG}_{S_V}$ entering into
cross sections. With the familiar Sudakov technique we find analogously to the
purely virtual case:
\begin{eqnarray}
{\cal F}^{{RG}_{2L}}_{S_R}+ 2{\cal F}^{{RG}_{2L}}_{S_V} &=& 
\frac{C^2_F}{4 \pi^2} \int^1_0
\frac{d \beta_1}{\beta_1} \int^1_0 \frac{d \alpha_1}
{\alpha_1}  \int^1_0 \frac{d \beta_2}{\beta_2}
\int^1_0 \frac{d \alpha_2}{\alpha_2} 
\frac{\alpha_s(m^2)}{1+c \; \log \frac{s \alpha_1 \beta_1}{m^2}}
\frac{\alpha_s(m^2)}{1+c \; \log \frac{s \alpha_2 \beta_2}{m^2}}
\nonumber \\ 
&\times& \Theta \left( \alpha_1+\beta_1-\frac{2 l_c}
{\sqrt{s}} \right) \Theta \left( \alpha_2+\beta_2-\frac{2 l_c}
{\sqrt{s}} \right) \left\{ \Theta ( \alpha_1
- \frac{m^2}{s} \beta_1) 
\right. \nonumber \\ && \Theta (\beta_1- \frac{m^2}{s} \alpha_1)
\Theta ( \alpha_2 - \alpha_1)  
\Theta ( \alpha_2
- \frac{m^2}{s} \beta_2) \Theta (\beta_2- \frac{m^2}{s} \alpha_2
) \nonumber \\ &&
\Theta ( \beta_2-\beta_1) + \Theta ( \alpha_1
- \frac{m^2}{s} \beta_1) \Theta (\beta_1- \frac{m^2}{s} \alpha_2)
\Theta ( \alpha_2 - \alpha_1) \nonumber \\ &&
\left. \Theta ( \alpha_2- \frac{m^2}{s} \beta_1) 
\Theta (\beta_2- \frac{m^2}{s} \alpha_2) \Theta ( \beta_1-
\beta_2) \right\} \label{eq:RpVRG2L}
\end{eqnarray}
A numerical evaluation of Eq. \ref{eq:RpVRG2L}-which contains
a different ``running scale'' for each loop-is given in Fig. \ref{fig:cutexp} and
compared to half the square of the one loop result in Eq. \ref{eq:RpVRG}. It can
be seen that the renormalization group improved massive Sudakov form factor
exponentiates as expected.

\end{document}